%Paper: hep-ph/9305303
%From: PHADS%TWNAS886.BITNET@pucc.Princeton.EDU
%Date: Mon, 24 May 1993 11:57 PDT

\magnification\magstep1
\hsize5.45truein
\hoffset.8truein
\baselineskip23truept
\topskip10truept
\def\SE{\langle T_{\mu\nu}\rangle}

{
\nopagenumbers
\hfill IP-ASTP-19-93

\hfill May, 1993

\null
\vfill

\centerline{\bf The Asymptotic Form of the Energy Density of}
\centerline{\bf Weakly Interacting Particles}
\centerline{\bf in a Static, Spherical Geometry}
\vskip.5in

\centerline{\it Achilles D. Speliotopoulos}
\footnote{}{\rm Bitnet address: PHADS@TWNAS886}

\vskip.5cm

\centerline{\it Institute of Physics}
\centerline{\it Academia Sinica}
\centerline{\it Nankang, Taipei, Taiwan 11529}

\vfill

\centerline{\bf Abstract}

\vskip.25in

{
 \baselineskip26truept
 \leftskip2in
 \rightskip2in
 \noindent The asymptotic form of the energy density for a gas of
 particles surrounding a sphere of mass $M$ and radius $R$ is
 studied using Einstein's equations. It is shown that if the pressure
 of the gas $p$ varies linearly with the energy density $\rho$ for
 small $\rho$, then $\rho\sim 1/r^2$ for large $r$.
}
\vfill
\supereject
}
\pageno=2
In the Schwarzchild solution of Einstein's equations for a
static, spherical geometry $[1]$ the spacetime surrounding the sphere
of mass $M$ and radius $R$ is taken to be empty and free of
particles. The real world is certainly not so clean and simple,
however and a more realistic model would also include a gas of particles
which surround the sphere. With the addition of these particles the
question then becomes what affect, if any, they will have on the
geometry of the spacetime. At first glance this would seem to be an
impossible question to answer. Not only must an equation of state for
the particles be known, this equation of state must also have been
calculated in the presence of the very gravitational field that it itself
determines. What we shall find, however, is that due to the
restrictions that Einstein's equations themselves put on the energy
density of the particles, the asymptotic behavior of this spacetime
will be surprisingly simple to determine.

In this paper we shall study some of the asymptotic properties of a
static, spherical geometry filled with a gas of particles. To be
specific, the system we shall be considering consists of a sphere
with mass $M$ and radius $R$ which is surrounded by an arbitrary gas
of particles. We shall not need to specify the type of particles
surrounding the sphere, nor shall we need to fix their temperature or
density. All that we shall require is that they have an equation of
state which has certain properties. For convenience we have also
assumed that the sphere has not undergone complete gravitational
collapse into a black hole and that the system as a whole is in
thermodynamic equilibrium. Moreover, unlike the usual treatment of
matter in general relativity, we shall not a prior\`{\i} confine the
particles to be within a sphere of any fixed radius, but will instead
let the system itself determine how far into the spacetime the
particles will spread.

Although we have used the word ``gas'' to describe the particles, this
is only a matter of convenience. The particles are not only allowed
to interact with gravity, thereby determining the geometry of the
spacetime, but also with each other. Since the most stringent
constraint on the energy density, and thus the geometry of the
spacetime, comes not from the equation of state of the particles, but
rather from Einstein's equations themselves, all that we need require
is  that the equation of state satisfy a few physically reasonable
requirements. In the end we shall find that if the particles interact
among themselves to any significant extent, then the spacetime will
be asymptotically flat. The energy density of the particles will die
off faster than $1/r^2$ and we may charactorize the particles as
being confined within a sphere of a certain radius which will require
a specific choice of the equation of state to determine precisely. If
the particles are very weakly interacting, on the other hand, then we
shall find that the spacetime is not asymptotically flat. The energy
density will die off as $1/r^2$ for large $r$ and we cannot
charactorize the particles as being confined within a sphere of any
fixed radius. The affect of the particles on the geometry of the
spacetime is instead all pervasive.

We begin with the most general form of the metric for a static,
spherical geometry $[2]${}
$$
ds^2 = -f dt^2 + hdr^2 + r^2 d\theta^2 + r^2\sin^2\theta d\phi^2\>,
$$
where $f$ and $h$ are unknown functions of $r$ only. We are only
interested in the structure of the spacetime for $r>R$ and shall
always assume that this holds. As usual, we shall take the average
energy momentum tensor for the particles to be of the form $\langle
T_{\mu\nu}\rangle = \rho u_\mu u_\nu + p (g_{\mu\nu} + u_\mu u_\nu)$
where $u_\mu$ is a unit velocity vector which lies in the direction
of the timelike Killing vector for the system, while $\rho$ and $p$
are the energy density and pressure, respectively, of the particles
surrounding the sphere. The two Einstein's equations we shall find of
use are $[2]${}
$$
\eqalignno{
8\pi\rho =
          &
          {h'\over r h^2} + {1\over r^2}\left(1-{1\over h}\right)\>,
          &(1a)
          \cr
8\pi p =
          &
          {f'\over r fh} - {1\over r^2}\left(1-{1\over h}\right)\>,
          &(1b)
          \cr
}
$$
where the primes denote derivatives with respect to $r$ and we are
using units in which $G=c=1$.  From the conservation equation
$\nabla^\mu \SE = 0$ we also have the Tolman-Oppenheimer-Volkoff (TOV)
equation $[2]$ for hydrostatic equilibrium
$$
{dp\over dr} + {1\over 2}\left(p+\rho\right){1\over f}{df\over dr} = 0\>.
\eqno(2)
$$

At this point we shall assume that we are given an equation of state
for the particles which we shall write as
$$
p = w(\rho)\rho\>,
\eqno(3)
$$
where $w(\rho)$ is a functional of $\rho$. It is further assumed that
this equation of state includes the rest mass of the particles and
was calculated in the presence of the gravitational field. As such both
$p$ and $\rho$ must also be functions of $r$. We next make the anzatz
that $w(\rho)$ itself has no {\it explicit\/} $r$ dependence, but
is instead dependent on $r$ only through its dependence on $\rho$. We
justify this anzatz with the following. First, we know that in
Minkowski spacetime the equation of state may always be written in
the form given in eq.~$(3)$. As the system is in
thermodynamic equilibrium, and as there are no external fields
present, any $r$ dependence in $w(\rho)$ must then be due to the
curvature of the spacetime. Second, $\log f$ may, under certain
circumstances, be interpreted as being twice the gravitation
potential and is therefore  a measure of the ``gravitational force''
on the particles. Third, we note that all explicit $r$ dependence
in the TOV equation can be canceled and one can instead consider it
as a differential equation determining $p$ in terms of $f$. In fact,
any dependence of $p$ on $r$ is due to $f$. Finally, the geometry of
the spacetime, namely $f$ and $h$, is ultimately determined by the
energy density and pressure of the particles. Since the equation of
state on curved spacetimes must also reduce to the equation of state
calculated in Minkowski spacetimes in the limit where $f\to1$ and
$h\to1$, we would therefore expect $w(\rho)$ to depend on $r$ only
implicitly through $\rho$.

The equation of state will in general be very complicated, if it can
be calculated at all. Fortunately, we shall not need any specific
form of eq.~$(3)$, but rather that it have a couple of physically
reasonable properties which we would expect from any physical system.
First, the pressure $p$ must be finite as $\rho \to 0$, meaning that
the gas of particles cannot become infinitely stiff as the energy
density of the particles is reduced to zero. Actually, we shall need
the more stringent requirement that
$$
\lim_{\rho\to0} w(\rho)\equiv w_0\>,
$$
exists and is a finite. This means, in particular, that for small
$\rho$, $p\sim \rho^{1+n}$ for $n>0$. We justify this requirement
with the observation that if the particles were consisted only of
photons, then the pressure must be one third the energy density. This
holds even in curved spacetimes. Since the photons do not
interact among themselves, we have at least one example of a gas of
non-interacting particles for which the pressure varies linearly with
the energy density. We would therefore expect that if the particles do
interact with one another, then the pressure should vary at least
linearly with the energy density, if not stronger. We would not
expect the pressure to have a weaker power law dependence on the
energy density than a linear one.

With the Einstein's equations, and the TOV equation, we have now three
differential equations determining $\rho$, $f$, and $h$. (Since the
equation of state is given, $p$ is determined as soon as $\rho$ is.)
We shall then, presumeably, need a set of three initial conditions to
determine the system completely. The initial condition for $\rho$ is
straightforward; we need only take take its value at the
surface of the sphere $\rho_R\equiv \rho(R)$. The initial conditions
for $f$ and $h$, on the other hand, are much more difficult to
determine. Fortunately, we are only interested in the asymptotic
nature of the spacetime and we shall later find that the asymptotic
forms of both $\rho$ and $h$ are independent of their initial
conditions. Whatever initial condition we choose for $h$ is
thus immaterial for our purposes. $f$, on the other hand, {\it
does\/} depend on initial conditions; not only on its own, but also
on $\rho_R$ as well. Since, however, we may always rescale the time
coordinate, its dependence on its own initial condition is not
extremely relevant and we may formally take it to be $f(R)$.

Because the equation of state may be quite complicated,
the solution of eqs.~$(1)$ for all $r$ is not trivial
to find. In, however, the asymptotic limit where $r\to\infty$ the
situation simplifies dramatically. To see how this happens, we
integrate eq.~$(1a)$ to formally give
$$
{1\over h} = 1 - {2m_0\over r} - {8\pi\over r}\int^r_R \rho(\bar r)
{\bar r}^2 d\bar r\>,
$$
for $r>R$ where $m_0$ is an integration constant. Since $\rho$ is not
known explicitly, this would seem to be of little use. Note,
however, that $h>0$ for all $r>R$. Consequently, $\rho\to 0$ as
$r\to\infty$. In fact, $\rho$ must die off at least as fast as
$1/r^2$. Thus, because we are only interested in the asymptotic
behavior of the spacetime, only the behavior of the equation of state
when $\rho\to0$, namely $w_0$, will be relevant in our analysis.
Keeping this in mind, we next make use of the equation of state in
eq.~$(2)$ to obtain
$$
{1\over f}{df\over dr} = -{2\over 1+w(\rho)}\left(\rho{dw\over d\rho}
+ w(\rho)\right) {1\over \rho}{d\rho\over dr}.
\eqno(4)
$$
We then define
$$
\rho = {\Delta\over 4\pi r^2}\>,
$$
and note that $\Delta$ either vanishes as $r\to\infty$ or else
approaches a constant value. The important point is that $\Delta$
must be finite as $r\to\infty$. Then defining $h^{-1} = 1-2K$, eqs.
$(1)$ become
$$
\eqalignno{
\Delta =
        &
        {dK\over dy} + K\>,
        &(5a)
        \cr
\Delta w_0
        =
        &
        -
        {w_0(1-2K) \over 1+w_0}
                \left(
                {1\over \Delta}{d\Delta\over dy} -2
        \right) - K\>,
        &(5b)
        \cr
}
$$
for large $y$ where $y\equiv\log(r/r_0)$ for some $r_0$. Solutions to
these equations have two different behaviors depending upon
whether or not $w_0$ vanishes.

\noindent{\it Case 1: $w_0 = 0$}

{}From eq.~$(5b)$, $K\to 0$ as $r\to\infty$. Consequently from
eq.~$(5a)$, $\Delta\to 0$ as well. This means that the energy density,
and consequently the pressure, must decrease faster than $1/r^2$
for large $r$. Then from eq.~$(4)$, $f\to c_k$ where $c_k$ is a constant
which can be set to $1$ by suitably rescaling the time coordinate.
Consequently, if $w_0=0$, then $f\to 1$ and $h\to 1$ as $r\to\infty$
and the spacetime is asymptotically flat.

\noindent{\it Case 2: $w_0\ne0$}

This case is much more interesting. We write eqs.~$(5)$ as a set of two
coupled, non-linear differential equations
$$
\eqalignno{
{dK\over dy} =
        &
        \Delta - K\>,
        &(6a)
        \cr
{d\Delta\over dy}
        =
        &
        - {1+w_0\over w_0}{\Delta \over 1-2K}
        \left\{
                \Delta w_0 + \left(1+5w_0\over 1+w_0\right) K
                -{2w_0\over 1+w_0}
        \right\}.
        &(6b)
}
$$
Fortunately, these two equations are autonomous, meaning they have no
explicit $y$ dependence. Obtaining asymptotic solutions to eqs.~$(6)$
is then straightforward and involves looking for fixed points $(\Delta_a,
K_a)$ of eqs.~$(6)$ where the derivatives of both $\Delta$ and $K$
vanish. (See $[3]$. We caution the reader that, depending on the
literature, the term ``critical point'' is often used instead of
``fixed point''.) From eqs.~$(6)$ one of these fixed points occurs at
$\Delta_a=0=K_a$, which is the $w_0=0$ case once again. A second
fixed point occurs at
$$
\Delta_a = K_a = {2w_0\over (1+w_0)^2 + 4w_0}.
$$
If we next expand eqs.~$(6)$ about this fixed point, then
$$
{d\>\>\>\over dy}\pmatrix{\delta\Delta\cr\delta K\cr}
        =
        \pmatrix{-{2w_0\over 1+w_0}&-{2(1+5w_0)\over (1+w_0)^2}\cr
                1&-1\cr
                }
        \pmatrix{\delta\Delta\cr\delta K\cr}\>,
\eqno(7)
$$
where $\delta \Delta = \Delta-\Delta_a$ and $\delta K = K-K_a$.
The asymptotic behavior of the solutions to eqs.~$(6)$ depends on the
eigenvalues $\lambda_\pm$ of this matrix. Writting $\lambda_\pm=
-\eta\pm i\varphi$, we find that
$$
\eta = {1\over 2} \left(1+{2w_0\over 1+w_0}\right)\>,\qquad\qquad
\varphi = {1\over 2} {(7+42w_0-w_0^2)^{1/2}\over 1+w_0}.
$$
Since $\rho\ge0$ for all $r$, $\Delta_a$ must be positive. This
condition, combined with the requirement that $h^{-1} = 1-2K_a >0$
for all $r$, means that $w_0>0$. Moreover, because $p\le \rho/3$,
$w_0\le 1/3$, and it is then straightforward to see that $\varphi$ will
always be a real number. Consequently, this fixed point is {\it
stable\/} and all solutions to eqs.~$(6)$ will eventually spiral
counterclockwise into this fixed point in the $\Delta$-$K$ plane no
matter what their initial conditions were originally. We can see this
explicitly by solving eq.~$(7)$ to give
$$
\eqalign{
K(r) \approx
        &
        \Delta_a
        \left\{1 + A
                \left(
                        r\over r_0
                \right)^{-\eta}
                \sin{
                        \left(
                                \varphi \log{{r\over r_0}}
                        \right)
                }
        \right\}\>,
        \cr
\Delta(r) \approx
        &
        \Delta_a
        \Bigg\{
                1 + A
                \left(
                        r\over r_0
                \right)^{-\eta}
                \Bigg[
                        \varphi\cos
                        \left(
                                \varphi\log{r\over r_0}
                        \right)
        \cr
        &
                        +
                        (1-\eta)\sin
                        \left(
                                \varphi\log{r\over r_0}
                        \right)
                        \Bigg]
        \Bigg\},
        \cr
}
$$
for large $r$.  $A$, and $r_0$ are constants which depend on the
specific equation of state and initial condition for $\rho$ and $p$.
Notice also that because $0<w_0\le1/3$, $0<\Delta_a\le 3/14$, and
$1/2<\eta\le3/4$ while $\sqrt7/2<\varphi\le\sqrt{47}/4$. Solutions to
eqs.~$(6)$ therefore approaches this fixed point very slowly; the
fastest being $r^{-3/4}$.

{}From eq.~$(4)$ we can solve for $f$ in terms of $\rho$. In the
asymptotic limit we find that $f \approx k r^q$ where $q =
4w_0/(1+w_0)$,
$$
\eqalignno{
k=
        &
        f(R)
        \left(4\pi\rho_R\over\Delta_a\right)^{q/2}
        \exp\Bigg\{
        2\left(\rho_R {dw\over d\rho}\vert_{\rho_R}+w_R
                \over
                1+w_R
        \right)
        -
        {q\over 2}
        \cr
        &
        +
        2\int_0^1 x\log{x}{d^2\>\>\over
        dx^2}\left(xw'(x)+w(x)\over 1+w(x)\right)\>dx
        \Bigg\}
        \cr
}
$$
and $w_R\equiv w(\rho_R)$. Since $0<w_0\le 1/3$, $0<q\le 1$.
Consequently, for large $r$ the metric becomes
$$
ds^2 = -kr^q dt^2 + {1\over 1-2K_a}dr^2 + r^2 d\theta^2 + r^2\sin^2\theta
d\phi^2\>,
$$
and we can see explicitly that this spacetime is {\it not\/}
asymptotically flat.

To summerize, we have shown that if the particles interact among
themselves to such an extent that $w_0=0$, then a static,
spherical spacetime containing these particles is
asymptotically flat, as expected. Moreover, we see that the energy
density of these particles must necessarily decrease faster than $1/r^2$;
most probably very much faster. As the Schwarzchild geometry is also
asymptotically flat and involves a mass density which is confined
within a sphere of definite radius, we may therefore charactorize
these particles as effectively being confined within a sphere of a
certain radius which will require a specific equation of state to
determine. The affect of these particles on the curvature of the
spacetime, like the affect of the mass $M$ in the Schwarzchild
solution, eventually dies away and the spacetime asymptotically
approaches the Minkowski spacetime. When, on the other hand, the
particles interact with each other so weakly that $w_0\ne0$, then
their energy density decreases as $1/r^2$ and their affect on the
curvature of the spacetime is correspondingly long range. They
cannot be charactorized as being contained within a sphere of any
definite radius and are instead spread throughout the spacetime. In
fact, we find that $f\sim r^q$ while $h$ approaches a constant as
$r\to\infty$ and a spacetime filled with these weakly interacting
particles is not asymptotically flat.

What is of even greater interest is the universal nature of the
energy density in the asymptotic limit
$$
\rho_a = {\Delta_a c^4\over 4\pi r^2 G}\>,
\eqno(8)
$$
where we have replaced the correct factors of $c$ and $G$. (The
subscript $a$ denotes the asymptotic limit.) Notice
that the form of $\rho_a$ is the same irrespective of the mass
and radius of the sphere, irrespective of any ``initial condition''
for $\rho$ at $r=R$, and irrespective of any form that the equation
of state may take so long as $w_0\ne0$. In fact, the only property of
the particles that it does depend upon is $w_0$, a dimensionless
number. From dimensional arguments, if the particles surrounding the
sphere are massless, then $w_0$ must simply be a number. It cannot
even depend on the temperature of the gas, since, aside from the
Planck mass, there is no other length scale one can use to construct
a $w_0$. If, on the other hand, the particles have a mass $m$, then
$w_0$ is either once again a number, or else is a function of the
ratio $T/m$, where $T$ is the temperature of the gas.

Notice that although this universality extends to $h$, it does not
extend to include $f$. For large $r$, $f\approx kr^q$ and although
$q$ depends only on $w_0$, $k$ depends not only on $f(R)$ but also on
$\rho_R$. Consequently, the motion of test particles in this
spacetime will always be dependent the specific choice of initial
conditions for $f$ and $\rho$, and thus on the detail properties of
the particles and of the spherical mass.

The question remains as to how one may go about calculating $w_0$.
For certain cases this is trivial. Suppose $T_{\mu\nu}$
is the energy momentum operator for the particles such that its
equilibrium average is $\SE = \rho u_\mu u_\nu + p(g_{\mu\nu}+u_\mu
u_\nu)$. If this energy momentum operator is traceless, as it is for
a gas of pure photons, then $\langle T_\mu^\mu\rangle =0$ and $p =
\rho/3$. Consequently, any system which has a traceless energy
momentum operator has a $w(\rho) = w_0 = 1/3$. Calculating $w_0$ for
other systems, on the other hand, is a much more formidable task,
although at first glance not an impossible one. Because $w_0$ is
determined when the energy density of the particles is vanishingly
small, this suggests that when calculating $w_0$ one may, as a first
approximation, neglect the affect of the particles on the curvature
of the spacetime and may instead treat them as test particles. One
can then in principle use the vacuum solutions of Einstein's
equations as a background field and calculate the equation of state
for the system using methods described in $[4]$ .

We end this paper with a couple of observations. First, the $1/r^2$
behavior in eq.~$(8)$ is precisely what one would expect for the energy
density of ``dark matter'' based on Newtonian gravity $[5]-[7]$.
Unfortunately, when one has gone to such a large $r$ such that eq.~$(8)$
holds, one also finds that the spacetime no longer close to being
Minkowskian. It is instead extremely curved (see $[8]$). Second, the
form of $\rho_a$ is quite peculiar in that it contains no explicit
length scale for $r$. The only length scale which can be constructed
solely from universal constants, however, is the Planck length
$l^2_{pl} = \hbar G/c^3$. If we use it, then we can write eq.~$(8)$
as
$$
\rho_a=\rho_{pl}{\Delta_a\over 4\pi} \left(l_{pl}\over r\right)^2
\eqno(9)
$$
where $\rho_{pl}= c^7/(\hbar G^2)$ is the Planck energy density.
Although $\rho_{pl}$ is very large, eq.~$(9)$ is valid only at very large
$r$. Since $l_{pl}\sim 10^{-33}$cm, this ensures that the size of $\rho$
will always be physically reasonable.

\vfill

\centerline{\bf Acknowledgements}

This work is supported by the National Science Council of the
Republic of China under  contract number NSC 82-0208-M-001-086.
\vfill
\supereject
\centerline{\bf REFERENCES}
\item{$[1]$}K. Schwarzchild, Kl. Math.-Phys. Tech., 189-196 (1916),
            Kl. Math.-Phys. Tech., 424-434 (1916).
\par
\item{$[2]$}R. M. Wald, {\sl General Relativity}, Chapter 6, (The
                University of Chicago Press, Chicago, 1984).
\par
\item{$[3]$}C. M. Bender and S. A. Orszag, {\sl Advanced Mathematical
                Methods for Scientists and Engineers}, Chapter 4,
                section 4.4, (McGraw-Hill Book Company, New York, 1978).
\par
\item{$[4]$}N. D. Birrell and P. C. W. Davies, {\sl Quantum Fields in
                Curved Space}, Chapters 1,2 (Cambridge University
                Press, Cambridge 1982).
\par
\item{$[5]$}V. C. Rubin, W. K. Ford, and N. Thonnard, {\sl Ap. J.
            Letters}, {\bf 225} L107 (1978).
\par
\item{$[6]$}V. Trimble, {\sl Ann. Rev. Astron. and Astrophys.}, {\bf
            25} 425 (1987).
\par
\item{$[7]$}E. W. Kolb and M. S. Turner, {\sl The Early Universe},
            Chapter 1 (Addison-Wesley Publishing Company, Inc.,
            New York, 1990). E. W. Kolb and M. S. Turner, {\sl The
            Early Universe: Reprints}, Chapter 1 (Addison-Wesley
            Publishing Company, Inc., New York, 1988).
\par
\item{$[8]$}A. D. Speliotopoulos, submitted to {\sl Phys. Rev.
                Lett.}, April 1993.
\vfill

\end